\documentclass[]{llncs}
\usepackage{graphicx}
\usepackage{makecell}
\usepackage{tabularx}
\begin{document}
%
\title{Enhancing Document Retrieval in COVID-19 Research: Leveraging Large Language Models for Hidden Relation Extraction}

%
%
\author{Hoang-An Trieu$^1$ \and
Dinh-Truong Do$^1$ \and 
Chau Nguyen$^1$ \and
Vu Tran$^2$ \and
Minh Le Nguyen$^1$}

\authorrunning{Trieu et al.}
%
%

\institute{$^1$Japan Advanced Institute of Science and Technology, Ishikawa, Japan\\
\email{\{antrieu, truongdo, chau.nguyen, nguyenml\}@jaist.ac.jp}\\
$^2$The Institute of Statistical Mathematics, Japan\\
\email{vutran@ism.ac.jp}}
\maketitle              
\begin{abstract}
In recent years, with the appearance of the COVID-19 pandemic, numerous publications relevant to this disease have been issued. Because of the massive volume of publications, an efficient retrieval system is necessary to provide researchers with useful information if an unexpected pandemic happens so suddenly, like COVID-19. In this work, we present a method to help the retrieval system, the Covrelex-SE system, to provide more high-quality search results. We exploited the power of the large language models (LLMs) to extract the hidden relationships inside the unlabeled publication that cannot be found by the current parsing tools that the system is using. Since then, help the system to have more useful information during retrieval progress. 

\end{abstract}

\section{Introduction}
When a breakout such as COVID-19 occurs, relevant research increases daily with incredible speed. In the record of CORD-19023
dataset \cite{wang2020cord}, there are more than 900K papers introduced by March 31st, 2022. With that massive amount of scientific publications, a retrieval system that could provide useful documents plays a crucial role in the research process of exploring methods to cope with the pandemic. 
Using the Covrelex-SE as the baseline system,  we retrieve the relevant documents after receiving a query from the researcher by finding the documents that contain the relation that matches the input query. A proper relation is an object that involves three components ( \textit{arg1}, \textbf{relation}, \textit{arg2}) where \textit{arg1} and \textit{arg2} are the noun phrases that contain the entities in domains such as bio-medical, chemical, genetic and so on. Besides, \textit{relation} is an expression that describes the relationship between \textit{arg1} and \textit{arg2}.

A \textit{relation} can be found inside the scientific publications if those publications contain any sentences that describe this \textit{relation}. In \cite{tran2021covrelex}, the system they proposed applied parsing tools such as OpenIE to extract those \textit{relation} from the abstracts of the documents inside the CORD-19 dataset, and they show a positive result while taking this approach. However, this method could not afford to discover the hidden relation through the whole document. Therefore, we propose a method to generate the hidden relations in whole documents by exploiting the power in presenting language of LLMs such as Flan-T5 and Mixtral through prompting engineering. We found out that generating relations from the input documents gives us more interesting information when compared with the approach that heads to extract the relation in the documents. In summary, our contributions in this work are as follows: \textbf{(I)} proposing a method to harness LLMs' text generation abilities to reveal hidden relations within unlabeled documents, and \textbf{(II)} conducting a manual evaluation task to validate the effectiveness of our proposed approach.

\section{Background}
Since the COVID-19 outbreak appeared, collecting scientific information from many relevant publications has received much attention. Tran et al. \cite{tran2021covrelex} introduced CovRelex, a retrieval system for scientific publications that target entities and relations via relation extraction (RE) from COVID-19 scientific papers. Later, Do et al.\cite{do2023covrelex} proposed the Covrelex-SE system, which incorporated semantic information in relation-retrieving progress through text embedding. Zhang et al.\cite{zhang2020covidex} released Covidex, a search engine that allows users to check the information inside the CORD-19 dataset. Sohrab et al.\cite{sohrab-etal-2020-bennerd} introduced the BENNERD system, which captures the named entities in the biomedical domain and structures them under the unified medical language system (UMLS) for further research.

For RE tasks, the approach that treats the RE problem as the sequence generating has received the attention of researchers recently. Yang et al.\cite{yang2022fpc} proposed a fine-tuning method that applies prompting engineering to control the process of generating the relation of entities inside the documents. The performance in RE of LLMs is investigated by Wadhwa et al.\cite{wadhwa2023revisiting}, and they report a very promising result when using the GPT-3 model on the RE task. However, applied relation generating in specific domains such as biomedical without annotated documents is still a challenging task because of the risk that LLMs may not have enough insight knowledge to represent the full meaning of the input sequence and the shortage of data that causes difficulty in providing enough samples for the LLMs to adapt with the domain. This challenge motivated us to develop a system that can exploit the power of LLM to extract the relation inside the unlabeled corpus.  

\section{Methods}
\subsection{Overview}

\begin{figure}[htp]
    \centering
    \includegraphics[width=12cm]{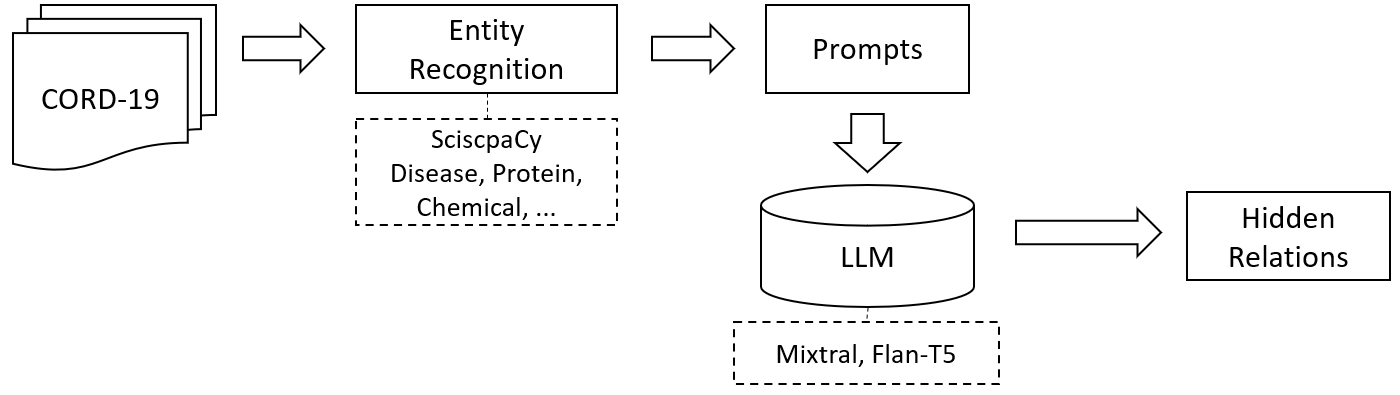}
    \caption{Overview of our system}
    \label{fig:overview}
\end{figure}

Figure \ref{fig:overview} illustrates an overview of our proposed method. Overall, our system first captures the named entities inside the documents from the CORD-19 dataset using the ScispaCy models\protect\footnotemark\hspace{0em}\footnotetext{\url{https://spacy.io/}}. Then we compare those entities with the database of the Covrelex-SE system \cite{do2023covrelex} to collect the pairs of entities that did not have any relations between them. After the set of entity pairs for each document is collected, we generate the prompt from the documents by marking the position of captured entities to instruct LLM on how to generate the relation between those entities. The results generated by LLM later is processed and stored in the database of the Covrelex-SE system to provide information for retrieving progress.

\subsection{Prompt Generating}
\label{prompt}
After extracting the entities from the documents, our target is to generate the relation between those entities inside the documents. However, meticulously examining every possible pair of entities from the entity pool would be inefficient in terms of time and resources. We assume that hidden relationships may exist between entity pairs that do not co-occur within the same sentence in the document. Therefore, we collected pairs of entities that appear in separate sentences and the sentences between them as the context to generate hidden relations. 
\begin{table}[!htbp]
    \centering
    \caption{Example for prompt generating}
    \begin{tabularx}{\linewidth}{|X X|}
    \hline
      \multicolumn{1}{|c|}{Original sentences}  &    Zika virus (ZIKV) is a flavivirus transmitted via mosquitoes and sex to cause congenital neurodevelopmental defects, including microcephaly... \\
    \hline
     \multicolumn{1}{|c|}{Head entity}  &   Zika virus  \\
     \hline
     \multicolumn{1}{|c|}{Tail entity}  &    flavivirus   \\
     \hline
     \multicolumn{1}{|c|}{Prompt 1}  &    extract all the relations inside the sentence in the sentence: " [E]Zika virus[/E] (ZIKV) is a [E]flavivirus[/E] transmitted via mosquitoes and sex to cause congenital neurodevelopmental defects, including microcephaly..."  that involves both "Zika virus" and "flavivirus". The results should be in JSON format that each JSON object has 3 keys which are "relation" and "entity1" and "entity2".   \\
     \hline
     \multicolumn{1}{|c|}{Prompt 2}  &  Find every relations between "Zika virus" and "flavivirus" in the sentence: " [E]Zika virus[/E] (ZIKV) is a [E]flavivirus[/E] transmitted via ..." The results should be in JSON format.   \\
     \hline
    \end{tabularx}
    
    \label{tab:prompt_example}
\end{table}
To generate the prompt to extract the relation between 2 entities inside a document using LLM, we inherited the idea of using the entity markers proposed by Yang et al. \cite{yang2022fpc}, but we removed the information of showing the head entity and tail entity inside the markers and let the LLM decide which entity should be head entity and vice versa. We use special tokens such as [E] and [/E] to mark the position of the entities that need to extract relations inside the input sentences. After that, we wrap the sentences with prepared templates to get the prompts for the LLMs. The templates we used are designed for Zero-shot settings. The example of the prompt template is illustrated in Table \ref{tab:prompt_example}.
\section{Experiments}
\subsection{Experimental Settings}



In this study, we aim to explore whether LLMs can detect hidden relationships between entities across different sentences. To achieve this, we selected 50 pairs of entities from a subset of 20 articles within the CORD-19 corpus. These entity pairs were then fed into the LLMs using a specific prompt (detailed in Section \ref{prompt}). The LLMs subsequently generated predictions regarding the relationship between each pair of entities. Following this, two evaluators independently assessed the accuracy of the predictions. They determined whether the identified relationships were correct, specifically focusing on hidden relationships not explicitly stated in the text but implied between the entities. The evaluators' judgments were then compared, and only predictions agreed upon by both evaluators were considered correct. Additionally, Cohen’s kappa coefficient \cite{mchugh2012interrater} was utilized to measure the level of agreement between the evaluators.

For the experimental setup, we employed two distinct LLM models: Flan-T5 \cite{chung2022scaling}, an encoder-decoder model with 11B parameters, and Mixtral-8x7B-Instruct-v0.1 \cite{jiang2024mixtral}, a decode-only model with 56B parameters. We utilized the Transformers library\protect\footnotemark\hspace{0em}\footnotetext{\url{https://huggingface.co/docs/transformers}} to handle the loading and inference processes for these models. Regarding hyperparameter settings during inference, we retained default values, with the temperature set to 1.0, top-p set to 1.0, and top-k set to 50.



\subsection{Main Results}

\setlength{\tabcolsep}{0.5em}
\renewcommand{\arraystretch}{1.3}
\begin{table}[ht]
\centering
    \caption{Results on the Multiple Sentence setting. }\label{tab:mul}
\begin{tabular}{|c|c|c|c|c|c|}
\hline
        & \textbf{Total} & \textbf{Correct I} & \textbf{Correct II} & \textbf{Correct I \& II} & \textbf{Kappa score} \\ \hline
Flan-T5 & 50    & 5 (10\%)  & 4 (8\%)    & 3 (12\%)        & 0.63        \\ \hline
Mixtral & 50    & 20 (40\%) & 18 (36\%)  & 15 (30\%)       & 0.66        \\ \hline
\end{tabular}
\end{table}

The evaluation results of our method are presented in Table \ref{tab:mul}. Correct I and Correct II are the number of cases that annotators I and II consider that be true. We observed that our approach utilizing the Mixtral model achieved 15 correct answers out of 50 queries (30\%). This outcome is promising as uncovering hidden relations between entities across sentences is challenging, and previous methods like OpenIE \cite{angeli2015leveraging} can only extract relations within a single sentence. In addition, the Mixtral model exhibits a significant advantage over the Flan-T5 model by 18\% in the task of generating relations. Several factors can contribute to these results: Firstly, the Mixtral model boasts a much larger size compared to the Flan-T5, with 56B parameters versus 11B. Secondly, the Mixtral model is newer and trained on more robust instruction data, enabling it to better understand human instructions. Furthermore, the Kappa scores in both settings exceed 0.6, indicating a good level of agreement \cite{fleiss2013statistical}.

\subsection{Case Analysis}
In this section, we present some analysis of several samples which involve positive cases and negative cases. Firstly, for the positive case shown in Table \ref{tab:good_sample}, the relation between "Coronavirus" and "Ribavirin" is mentioned in this paragraph, which is the "Combination of Ribavirin and Interferon-a" is "recommended to be a treatment" for the "The New Coronavirus Pneumonia". However, this relationship is not mentioned directly in any single sentence, so with an extraction tool such as OpenIE, we could not achieve this type of relation. When using the Mixtral model to generate the relation in this paragraph, we achieved the relation "treatment" between those phrases, which is very close to the actual hidden relation inside this paragraph. On the other hand, for the negative sample shown in Table \ref{tab:bad_sample}, we can see two main errors that reduce the performance of LLMs, the first one is that generated relations do not contain the target entities, the second one is that the hidden relation is not correct according to the context paragraph. Besides, we observed that the number of errors in that relations did not contain target entities occurred with higher frequency when compared with the other one, so for future work, assuring LLMs generate relations containing correct entities is an important task to improve the system. 
\begin{table}[!htbp]
    \centering
    
    \caption{Example of positive result}
    \begin{tabularx}{\linewidth}{|X X|}
         \hline
         \multicolumn{1}{|c|}{Input Sequence}  &  The New Coronavirus Pneumonia (NCP, also named as COVID-19 by WHO on Feb 11 2020, is now causing a severe public health emergency in China since. The number of diagnosed cases is more than 40,000 until the submission of this manuscript. Coronavirus has caused several epidemic situations world widely, but the present contagious disease caused by 2019 new Coronavirus is unprecedentedly fulminating. The published cohorts of 2019 new Coronavirus (n-Cov) are single-center studies, or retrospective studies. We here share the therapeutic experiences of NCP treatment with literature review. Combination of Ribavirin and Interferon-a is recommended by the 5(th) edition National Health Commission's Regimen (Revised Edition) because of the effect on MERS (Middle East Respiratory Syndrome), and the effectiveness of Lopinavir/Ritonavir and Remdisivir needs to be confirmed by randomized controlled trial (RCT), given the situation of no specific antivirus drug on NCP is unavailable.\\
         \hline
         \multicolumn{1}{|c|}{head entity}  & \multicolumn{1}{l|}{Coronavirus } \\
         \hline
         \multicolumn{1}{|c|}{tail entity}  & \multicolumn{1}{l|}{ Ribavirin} \\
         \hline
         \multicolumn{1}{|c|}{relation}  & [\{'relation': 'treatment', 'entity1': 'combination of Ribavirin and Interferon', 'entity2': '2019 new Coronavirus (n-Cov)'\}] \\
         \hline
    \end{tabularx}
    \label{tab:good_sample}
\end{table}

\begin{table}[!htbp]
    \centering
    \caption{Example of negative result}
    \begin{tabularx}{\linewidth}{|X X|}
         \hline
         \multicolumn{1}{|c|}{Input Sequence}  &  Escherichia coli O157:H7, an important food-borne pathogen, has become a major public health concern worldwide. The aim of this study was to investigate the molecular epidemiologic feature of E. coli O157:H7 strains in China. 105 E. coli O157:H7 isolates were collected from various hosts and places over 9 years. A multilocus sequence typing scheme (MLST) was applied for bacteria genotyping and polymerase chain reaction (PCR) was used for virulence factor identification.\\
         \hline
         \multicolumn{1}{|c|}{head entity}  & \multicolumn{1}{l|}{multilocus sequence  } \\
         \hline
         \multicolumn{1}{|c|}{tail entity}  & \multicolumn{1}{l|}{ coli} \\
         \hline
         \multicolumn{1}{|c|}{relation}  & [\{'relation': 'uses', 'head entity': 'multilocus sequence typing scheme (MLST)', 'tail entity': 'Escherichia coli O157:H7 strains'\}, \{'relation': 'applied for', 'head entity': 'multilocus sequence typing scheme (MLST)', 'tail entity': 'bacteria genotyping'\}, \{'relation': 'used for', 'head entity': 'polymerase chain reaction (PCR)', 'tail entity': 'virulence factor identification'\}, \{'relation': 'identified', 'head entity': 'Escherichia coli O157:H7', 'tail entity': '105 isolates'\}, \{'relation': 'collected', 'head entity': '105 E. coli O157:H7 isolates', 'tail entity': 'various hosts and places over 9 years'\}] \\
         \hline
    \end{tabularx}
    \label{tab:bad_sample}
\end{table}




\section{Conclusion}
In summary, our work opened a new improvement for the Covrelex-SE system, allowing the system to collect more useful information to have better accuracy in retrieving progress. Although there are drawbacks in our system when the performance in generating correct relations is not too high, the result also shows promising potential for future improvement. The instability in the generated sequences causes difficulty in our work to have a trusty performance. Solving this problem would help the system become more practical and help us to prepare for breakouts such as COVID-19 in the future. 


\bibliographystyle{splncs04}
\bibliography{ref}

\begin{thebibliography}{10}
\providecommand{\url}[1]{\texttt{#1}}
\providecommand{\urlprefix}{URL }
\providecommand{\doi}[1]{https://doi.org/#1}

\bibitem{angeli2015leveraging}
Angeli, G., Johnson~Premkumar, M.J., Manning, C.D.: Leveraging linguistic structure for open domain information extraction. In: Proceedings of the 53rd Annual Meeting of the Association for Computational Linguistics and the 7th International Joint Conference on Natural Language Processing (Volume 1: Long Papers). pp. 344--354. Association for Computational Linguistics, Beijing, China (Jul 2015). \doi{10.3115/v1/P15-1034}, \url{https://aclanthology.org/P15-1034}

\bibitem{chung2022scaling}
Chung, H.W., Hou, L., Longpre, S., Zoph, B., Tay, Y., Fedus, W., Li, Y., Wang, X., Dehghani, M., Brahma, S., et~al.: Scaling instruction-finetuned language models. arXiv preprint arXiv:2210.11416  (2022)

\bibitem{do2023covrelex}
Do, T., Nguyen, C., Tran, V., Satoh, K., Matsumoto, Y., Nguyen, M.: Covrelex-se: Adding semantic information for relation search via sequence embedding. In: Proceedings of the 17th Conference of the European Chapter of the Association for Computational Linguistics: System Demonstrations. pp. 35--42 (2023)

\bibitem{fleiss2013statistical}
Fleiss, J.L., Levin, B., Paik, M.C.: Statistical methods for rates and proportions. john wiley \& sons (2013), \url{https://books.google.co.jp/books?id=9VefO7a8GeAC}

\bibitem{jiang2024mixtral}
Jiang, A.Q., Sablayrolles, A., Roux, A., Mensch, A., Savary, B., Bamford, C., Chaplot, D.S., Casas, D.d.l., Hanna, E.B., Bressand, F., et~al.: Mixtral of experts. arXiv preprint arXiv:2401.04088  (2024)

\bibitem{mchugh2012interrater}
McHugh, M.L.: Interrater reliability: the kappa statistic. Biochemia medica  \textbf{22}(3),  276--282 (2012), \url{https://hrcak.srce.hr/89395}

\bibitem{sohrab-etal-2020-bennerd}
Sohrab, M.G., Duong, K., Miwa, M., Topi{\'c}, G., Masami, I., Hiroya, T.: {BENNERD}: A neural named entity linking system for {COVID}-19. In: Liu, Q., Schlangen, D. (eds.) Proceedings of the 2020 Conference on Empirical Methods in Natural Language Processing: System Demonstrations. pp. 182--188. Association for Computational Linguistics, Online (Oct 2020). \doi{10.18653/v1/2020.emnlp-demos.24}, \url{https://aclanthology.org/2020.emnlp-demos.24}

\bibitem{tran2021covrelex}
Tran, V., Tran, V.H., Nguyen, P., Nguyen, C., Satoh, K., Matsumoto, Y., Nguyen, M.: Covrelex: A covid-19 retrieval system with relation extraction. In: Proceedings of the 16th conference of the european chapter of the association for computational linguistics: system demonstrations. pp. 24--31 (2021)

\bibitem{wadhwa2023revisiting}
Wadhwa, S., Amir, S., Wallace, B.C.: Revisiting relation extraction in the era of large language models. arXiv preprint arXiv:2305.05003  (2023)

\bibitem{wang2020cord}
Wang, L.L., Lo, K., Chandrasekhar, Y., Reas, R., Yang, J., Burdick, D., Eide, D., Funk, K., Katsis, Y., Kinney, R., et~al.: Cord-19: The covid-19 open research dataset. ArXiv  (2020)

\bibitem{yang2022fpc}
Yang, S., Song, D.: Fpc: Fine-tuning with prompt curriculum for relation extraction. In: Proceedings of the 2nd Conference of the Asia-Pacific Chapter of the Association for Computational Linguistics and the 12th International Joint Conference on Natural Language Processing. pp. 1065--1077 (2022)

\bibitem{zhang2020covidex}
Zhang, E., Gupta, N., Tang, R., Han, X., Pradeep, R., Lu, K., Zhang, Y., Nogueira, R., Cho, K., Fang, H., et~al.: Covidex: Neural ranking models and keyword search infrastructure for the covid-19 open research dataset (preprint)  (2020)

\end{thebibliography}
\end{document}